\documentclass[12pt,preprint]{aastex}
\newcommand{\bm}[1]{\mbox{\boldmath$#1$}}
\newcommand{\simgt}{\lower.5ex\hbox{$\; \buildrel > \over \sim \;$}}
\newcommand{\simlt}{\lower.5ex\hbox{$\; \buildrel < \over \sim \;$}}

\newcommand{\baredth}{\;\overline{\raise1.0pt\hbox{$'$}\hskip-6pt
\partial}\;}
\newcommand{\edth}{\;\raise1.0pt\hbox{$'$}\hskip-6pt\partial\;}
\begin{document}
\title{Probing Intracluster Magnetic Fields with Cosmic Microwave
Background Polarization} \author{Hiroshi Ohno\altaffilmark{1,2},
Masahiro Takada\altaffilmark{3}, Klaus Dolag\altaffilmark{4}, Matthias
Bartelmann\altaffilmark{4},\\ and Naoshi Sugiyama\altaffilmark{1}}

\altaffiltext{1}{
Division of Theoretical Astrophysics,
National Astronomical Observatory,
2-21-1 Osawa, Mitaka, Tokyo 181-8588, Japan 
; ohno@th.nao.ac.jp, naoshi@th.nao.ac.jp }
\altaffiltext{2}{Research Center for the Early Universe, 
School of Science, University of Tokyo, Tokyo 113-0033, Japan
; ono@resceu.s.u-tokyo.ac.jp}
\altaffiltext{3}{
Department of Physics and Astronomy, University of Pennsylvania, 
209 S. 33rd Street, Philadelphia, PA 19104, USA
; mtakada@hep.upenn.edu}
\altaffiltext{4}{
Max-Planck-Institut f\"ur Astrophysik, P.O. Box 1317, D-85741, Garching, 
Germany}

%\affil{ono@resceu.s.u-tokyo.ac.jp;mtakada@th.nao.ac.jp; naoshi@th.nao.ac.jp}
\begin{abstract}
 Intracluster magnetic fields with $ \sim \mu \rm G $ strength
 induce Faraday rotation on the cosmic microwave background
 (CMB) polarization.  Measurements of this effect can
 potentially probe the detailed structure of intracluster magnetic
 fields across clusters, since the CMB polarization is a continuously
 varying field on the sky, in contrast to the conventional method
 restricted by the limited number of radio sources behind or inside a
 cluster.  We here construct a method for extracting information on
 magnetic fields from measurements of the effect, 
 combined with possible observations
 of the Sunyaev-Zel'dovich effect and $X$-ray emission for the same
 cluster which are needed to reconstruct the electron density fields.
 Employing the high-resolution magneto-hydrodynamic simulations
 performed by Dolag, Bartelmann \& Lesch (1999) as a realistic model of
 magnetized intracluster gas distribution, we demonstrate how our
 reconstruction technique can well reproduce the magnetic fields, i.e., 
 the spherically averaged radial profiles of the field strength and 
 the coherence length.
\end{abstract}
%%%%%%%%%%%%%%%%%%%%%%%%%%%%%%%%%%%%%%%%%%%%%%%%%%%%%%%%%%%%%%%%%%%%%%%%

\section{Introduction}
 The origin and evolution of cosmic magnetic fields are still unclear
 and outstanding problems.  Various observational techniques have
 consistently revealed that most clusters of galaxies are pervaded by
 magnetic fields of $B \sim \rm O(1) \mu \rm G$ strength (see e.g., Carilli
 \& Taylor 2002 for a review).  Recently, Clarke, Kronberg \&
 B\"ohringer (2001) have again drawn a robust conclusion that the
 intracluster hot plasma typically has a $B \sim 5-10 ~\mu$G field
 assuming a coherence length of $10 \rm{kpc}$ from the Faraday rotation
 measurements of 16  low-z ($ z \leq 0.1$) clusters
 selected to be free of unusual radio halos.  Since the rotation measure
 arises from the integration of the product of the electron density and
 the line-of-sight component of magnetic fields, it is generally
 difficult to extract information on the magnetic fields only from the
 rotation measure without introducing any assumptions on the gas
 distribution and the field configuration.

 The first systematic study of the structure of
 magnetic fields over a single cluster was performed by Kim et al. (1990)
 based on the rotation measures of 18 sources close in angular position
 to the Coma cluster, giving the estimation of $ B \sim 2\mu$G with the
 coherence length of $10 \rm{kpc}$ which is indicated from the magnetic
 field reversal scale.  In a subsequent study, Feretti et al. (1995)
 discovered smaller coherence lengths down to 1 kpc from the rotation
 measures of the extended radio galaxy near the Coma cluster center,
 whereby the field strength estimation was modified to $B \sim 8\mu \rm
 G$ to explain the measured rotation angle.  
 Thus, it is crucial for estimating the magnetic field strength
 robustly to determine the magnetic field coherence length which does not
 necessarily match the coherence length of the rotation measures.
 To study the detailed structure of the intracluster magnetic fields for
 any clusters, the high-resolution measurements of the Faraday rotation
 should at least  be performed.
 However, there are some
 limitations for the conventional methods  because
 of the lack of the number of radio sources behind or inside a cluster
 and possible contributions of the intrinsic Faraday rotation.
 Moreover, we should break the degeneracy of the rotation measure
 between the electron density and the magnetic field strength 
 using some additional information for estimating the magnetic field strength. 

 Recently, Takada, Ohno \& Sugiyama (2001) proposed that the magnetized
 intracluster gas similarly induces a Faraday rotation effect on the
 linearly polarized radiation of the cosmic microwave background (CMB)
 generated at the decoupling epoch of $z\approx 1000$ (see e.g., Hu \&
 White 1997 for a study of the primary CMB polarization).  
 Hereafter, to distinguish the origin of the CMB anisotropies, we use ``primary''
 to indicate the anisotropies from the recombination epoch while
 ``secondary'' to those from the cluster formation epoch.
 They also calculated
 the angular power spectra of 
 this secondarily induced polarization under the simple
 assumption of a uniform field strength with $\sim 0.1\mu~{\rm G}$ across
 a cluster and suggested that the measurements could be used to set
 constraints on the average properties of the intracluster magnetic
 fields.  Cooray, Melchiorri \& Silk (2002) computed the
 secondary power spectrum including a circularly polarized 
 contribution characterized by the Stokes-V
 parameter that is induced by possible relativistic plasma in clusters
 via the Faraday rotation. 
 However, as a more interesting and realistic
 possibility, one may imagine that 
 the secondary effect  
 can be in
 principle used to map the detailed structure of the magnetic fields in
 an individual cluster, since the CMB polarization is a continuously
 varying field on the sky.  In this paper, therefore, we study a method
 for reconstructing the magnetic fields from measurements of the Faraday
 rotation effect on the CMB polarization, combined with accessible
 observations of the Sunyaev-Zel'dovich (SZ) effect and $X$-ray emission
 for the same cluster to reconstruct the electron density distribution.
 For this purpose, it is crucial to consider a realistic magnetic field
 configuration as well as a plausible gas distribution as suggested by
 the currently favored formation scenario of galaxy clusters. Hence, as
 for a model of the magnetized cluster, we employ here high-resolution
 magneto-hydrodynamic simulation results performed by Dolag, Bartelmann \&
 Lesch (1999), whereby we can simulate maps of the CMB
 polarization including the Faraday rotation effect as well as maps of
 the SZ effect and the thermal $X$-ray emission. Using those simulated
 maps, we demonstrate how well our method can reconstruct the magnetic fields. 
 In particular, we try to clarify the relationship between the coherence lengths
 of the magnetic fields and the rotation measure.
%we pay special attention to developing the method
% of well reconstructing the coherence length of magnetic fields as a
% function of radius from the cluster center, which should give a new
% insight into the nature and evolutionary history of magnetic fields. 
 The coherence length should give a new insight into the nature and
 evolutionary history of magnetic fields.
 For example, if the magnetic fields have the coherence scale as large
 as or larger than the cluster size, 
 the seed fields should be 
 generated at the early stage of the universe (see e.g., Grasso
 and Rubinstein 2001 for a review),
 while smaller coherence lengths may imply the seed fields originated
 from galaxies within the cluster (e.g., Kronberg 1996).

 This paper is organized as follows. In \S 2, we refer to the cluster
 models which are used to demonstrate the magnetic field reconstruction.
 In \S 3, we briefly review the primary CMB polarization map generated
 at the decoupling epoch and the
 Faraday rotation effect.
 In \S 4, assuming spherical symmetry of clusters, 
 we show a method for reconstructing the density fields of 
 the clusters from SZ effect
 and thermal $X$-ray emission which are directly calculated from simulated clusters.
 In \S 5, we develop a way to reconstruct the coherence length 
 and strength of the
 magnetic fields combined with the previously reconstructed electron
 density fields.    
 Finally, \S 6 is devoted to a summary and discussion.

\section{Cluster models}
\begin{figure}[hp]
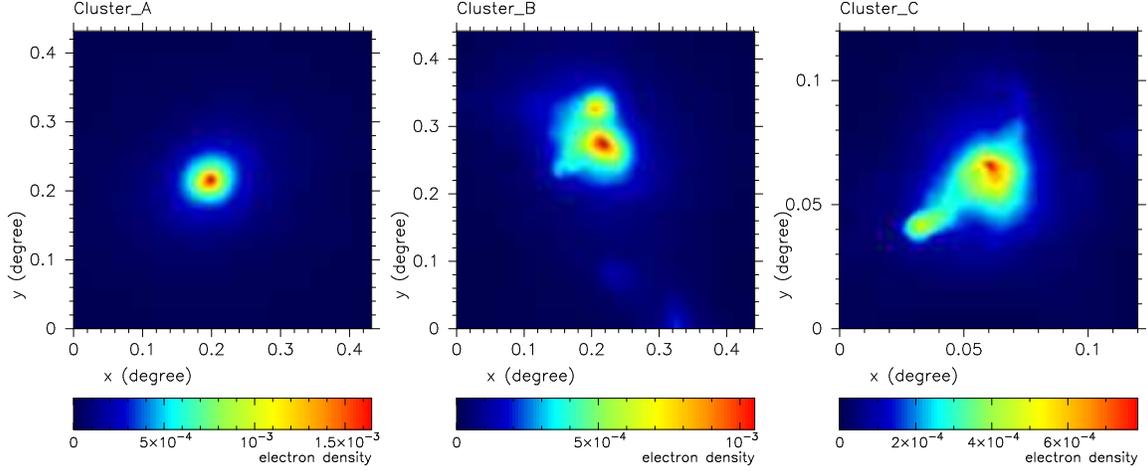

\epsscale{0.3}
\plotone{f1.eps} \plotone{f2.eps} \plotone{f3.eps}
\caption{Average electron density along the line of sight,
 $\langle n_e (1/cm^3) \rangle \equiv  1/(2L) \int_{-L}^{L} ds n_e$ for
 each cluster (A, B, and C). The axes  x and y 
 show the angular scale in degrees.} 
\label{fig:models}
\end{figure} 

\begin{deluxetable}{cccccc}
\small
 %\label{table:virial}
 \tablewidth{270pt}
\tablecaption{Cluster Models}
\tablehead{
& \colhead{$Cluster A$            }&  %\tablenotemark{b}}&
  \colhead{$Cluster B$            }&  %\tablenotemark{c}}& 
  \colhead{$Cluster C$            }& \\  %\tablenotemark{d}}& \\
}

\startdata
$z$                                             & 0.125         & 0.125
 & 0.452 \\
$r_{\rm {vir}}$ [Mpc] \tablenotemark{a}              & 2.7           & 2.7         & 2.1
 & \\
$M_{\rm {vir}}$      [$ \rm {M_\odot} $]  & $1.4 \times 10^{15}$  
& $1.3 \times 10^{15}$ & $1.1 \times 10^{15}$ & \\
$\Delta t$ [($\times 10^9$) years] \tablenotemark{b}
& 8&  2& 1&\\
\enddata

\tablenotetext {a}
{\footnotesize{ The centers of the clusters are taken at the electron density peaks.}}
%\tablenotetext{e}{\footnotesize{Mean Polarized Offset
%$\overline{T_{\textrm{pol}}}=\sqrt{\overline{Q}^2+\overline{U}^2}$,
%where $\overline{Q}$ and $\overline{U}$ are the Stokes parameter
%offsets. Numbers in parentheses denote the corresponding values
%for the QPC.}} \tablenotetext{f}
\tablenotetext{b}
{\footnotesize{Time interval since the epoch when
 the mass of the cluster had half of its final
 virial mass.}}
\label{table:virial}
\end{deluxetable}
 We employ high-resolution magneto-hydrodynamic 
 simulations performed by Dolag et al. (1999) as a realistic model of  
 magnetized clusters (see also Dolag et al. 2001a,b; 2002).
 The simulations start from the seed magnetic field with $10^{-9}$G
 strength at redshift $z=20$,    which corresponds  an
 upper limit to IGM fields set by Faraday rotation measurements of
 high-z radio loud QSOs (Kronberg 1996; Blasi, Burles, \& Olinto 1999).  
 They considered the homogeneous or chaotic field
 configurations for the initial seed fields,
 which are motivated by expectations of the primordial
 or galactic-wind induced initial seed fields, respectively. 
 Then, the evolutionary history of the magnetic field was followed 
 under the ideal magneto-hydrodynamic approximation. 
 Several interesting results were revealed in their paper. First, 
 the initial field configurations are not
 important for field configurations in the final stage of cluster evolution.
 Secondly, the final field strength is 
 amplified  to reach $1 \mu {\rm G} $ by
 gravitational collapse and shear flows which induce Kelvin-Helmholtz
 instabilities. 
 Although the simulation ignored effects associated with
 individual galaxies in the cluster on 
 the evolution of the magnetic field, it is worth mentioning
 that the simulation results can explain some observational implications of
 the rotation measure (also see Dolag et al. 2001a,b). 

% In the following, we consider three clusters which are the results of
% magnetic hydrodynamics cosmological simulations.
% The simulations are described in detail in two earlier papers
% (e.g. Dolag et al. 1999, 2002).
% Briefly, cosmological initial conditions including magnetic seed
% fields are set up at high redshift ($z=20$), and the collapse of high
% density regions into galaxy clusters is followed, consistently
% including the evolution of the magnetic fields.
 In the following, we consider three clusters for demonstrating
 the magnetic field reconstruction, and 
 hereafter we refer to these clusters as model A, B and C. 
 The redshifts $z$, virial radii $r_{\rm {vir}}$, virial masses
 $M_{\rm {vir}}$ and the interval time  
 since the formation of the clusters, $\Delta t$,
 are given in
 Table \ref{table:virial}.
 Here, we simply define the formation time as the epoch
 when the cluster had half of its final virial mass.
 Fig. {\ref{fig:models}} shows
 the average electron density $\langle n_e \rangle$ along the line of
 sight $s$, $\langle n_e \rangle \equiv  1/(2L) \int_{-L}^{L} ds n_e$,
 where $2L$ (= 4 Mpc)  denotes the size of the
 simulation volume.
 The electron density distribution of model A looks spherically
 symmetric. 
 Since the main amplification mechanism in
 the simulation is gravitational collapse, 
 the magnetic field strength should be roughly
 in proportion to the electron density.
 Thus, we expect that the distribution of magnetic field strength can be
 also roughly considered as spherically symmetric. 
 For this cluster model, therefore, the spherical symmetry approximation, which is a basic
 assumption for the following reconstructions, would be appropriate for 
 both the magnetic field strength and the electron density. 
 Model B has two density peaks in the cluster inner region, and the ratio of
 these peaks is $\sim $ 1 .  Hence, as a result of our analysis which 
 takes the electron density peak as the cluster center, 
 the reconstructed quantities  may deviate from the spherically averaged
 true values around the second peak. 
 Model C has a subcluster in the outskirts region, and the spherical 
 symmetry assumption may
 not be appropriate for this cluster around the subcluster. 
 Thus, the B and C clusters are considered to demonstrate that 
 our method can work well for clusters with substructures.   
 Note that the redshift of C cluster is higher
 than those of the other two clusters, 
 which is the one reason for the elongated 
 structure. The simulation sequence indeed shows that the subcluster is 
 merged with the main cluster component at the lower redshift.  
 It should be again stressed that we will take the cluster centers 
 as the electron density peaks in the following analysis. 
 
 The ratio $\beta$ between thermal and magnetic pressure in the
 simulated clusters is specifically addressed in Dolag et al. (2001a),
 where it is shown in Fig. 1 that the $\beta$ parameter is almost
 constant within the virial radius.

\section{Rotation measure and primary polarization}
\subsection{Small scale limit approximation of the Stokes parameters 
\label{sec:sec1}}
 Since the Faraday rotation effect due to the magnetized clusters 
 on the CMB polarization is important only on the small angular scales, 
 we can safely employ the small angle approximation \cite{ZS98,Takada}.
 In this limit, the $Q$ and
 $U$ fields of the Stokes parameters 
 can be expressed using the two-dimensional Fourier transformation
 in terms of the electric $(E)$ and magnetic $(B)$ modes (Kosowsky
 1996; Zaldarriaga,
 \& Seljak 1997;Kamionkowski, Kosowsky, \& Stebbins 1997 a,b; Hu, \&
 White 1997 a,b) as
\begin{eqnarray}
Q(\bm{\theta})&=&\int\!\!\frac{d^2\bm{l}}{(2\pi)^2}
\left[E_{\bm{l}}\cos2\phi_{\bm{l}}-
B_{\bm{l}}\sin2\phi_{\bm{l}}\right]e^{i\bm{l}\cdot\bm{\theta}},\nonumber\\
U(\bm{\theta})&=&\int\!\!\frac{d^2\bm{l}}{(2\pi)^2}
\left[E_{\bm{l}}\sin2\phi_{\bm{l}}+B_{\bm{l}}\cos2\phi_{\bm{l}}\right]e^{i\bm{l}
\cdot\bm{\theta}},
\end{eqnarray}
 where $\bm l$ and $\phi_{l}$ are defined as $\bm{l}\equiv
 l(\cos\phi_{\bm{l}},\sin\phi_{\bm{l}})$ 
 in the fixed (x,y)-orthogonal coordinates.
%\begin{eqnarray}
%\bm l    & \equiv & (l_x,l_y), \\
%l_x+il_y & \equiv & le^{i \phi_{{l}}}.
%\end{eqnarray}
 Here, $E_{\bm{l}}$ and $B_{\bm{l}}$ are the Fourier coefficients for
 the primary $E$-
 and $B$-modes, respectively.
 We use the CMBFAST code \cite{SZ96} to generate 
 $E_{\bm{l}}$, and ignore the $B$-mode for simplicity, 
 since the vector- and tensor-mode primordial fluctuations that induce
 the $B$-mode have not so far been detected.  
 Thus, under the assumption of Gaussian
 random fields, we can easily make a primary polarization map (e.g., 
 see a method of making the primary CMB temperature map in Takada \&
 Futamase 2001 in more detail). 

\subsection{Faraday rotation}
\begin{figure}[ht]
\epsscale{0.5}
\plotone{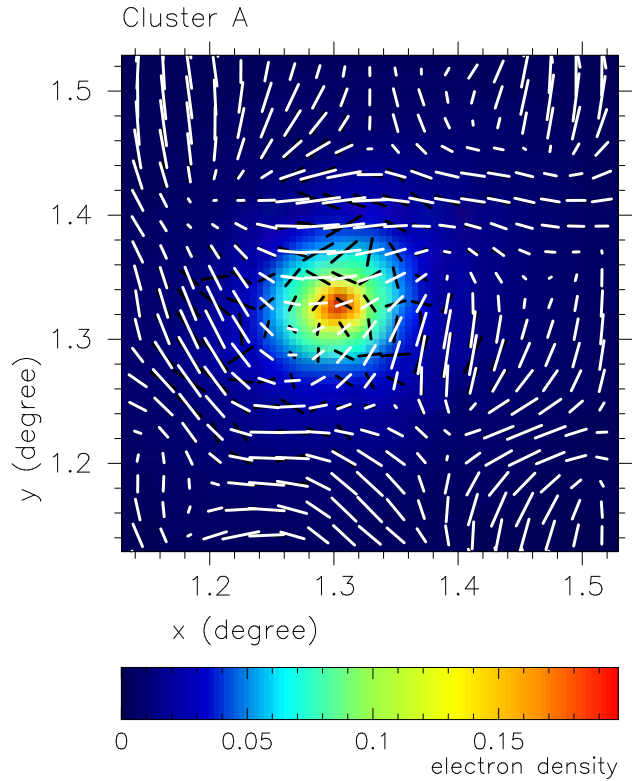}
\caption{A example of the simulated map of the  primary CMB 
 polarization (white lines) and the map (black lines) 
 including the Faraday rotation effect for cluster A model. 
The map is overlapped with  the average density fields along the line
 of sight, $\langle n_e ({\rm cm}^{-3}) 
\rangle \equiv  1/(2L) \int_{-L}^{L} ds n_e$ as in
 Fig. \ref{fig:models}.
 The axes x and y show the angular scale in degrees.}
\label{fig:polari}
\end{figure}

 If a linearly polarized monochromatic radiation of frequency $\nu$ 
 is passing through a
 hot plasma in the presence of magnetic fields along the
 propagation direction $\bm{s}$, 
 the polarization vector will
 be rotated by the angle (the rotation measure), $\Delta_{RM}$ (see
 e.g., Rybicki \& Lightman 1979);
\begin{eqnarray}
\Delta_{RM} &=& \frac{e^3}{2\pi m_e^2c^2\nu^2}
\int\!\!ds n_e(\bm{B}\cdot \hat {\bm{s}})\nonumber\\
&\approx &
8.12\times 10^{-2} (1+z)^{-2}\left(\frac{\lambda_0}{1{\rm ~cm}}\right)^{2}
\int\!\!\frac{ds}{{\rm ~kpc}}
\left(\frac{n_e}{1{\rm ~cm}^{-3}}\right)\left(\frac{\bm{B}\cdot \hat{\bm{s}}}
{1~\mu{\rm G}}\right),
\label{eqn:rm}
\end{eqnarray}
 where $\lambda_0$ is the observed wavelength, and $e$ and $m_e$ denote the
 electron charge and mass, respectively. 
 This Faraday rotation effect
 should rotate the primary CMB polarization vector as a secondary effect.
 From eq. (\ref{eqn:rm}), one can see that the rotation measure arises from
 the combined contributions of the electron density $n_e$ and the
 line-of-sight component of the magnetic
 fields; $B_s \equiv \bm{B}
 \cdot \hat{\bm{s}}$. To break this
 degeneracy, we first consider a way to reconstruct
  the electron density fields in the next section.
 It is worth noting that the wavelength dependence could be used to
 separate the Faraday rotation effect  from the primary CMB
 polarization and other secondary effects.
  Fig. \ref{fig:polari} shows an example of simulated CMB
 polarization maps with and without the secondary effect due to the
 Faraday rotation effect, where we have considered cluster A model. 

%%%%%%%%%%%%%%%%%%%%%%%%%%%%%%%%%%%%%%%%%%%%%%%%%%%%%%%%%%%%%%%%%%%%%
\section{Reconstruction of $n_e$ \label{sec:nere}}
 To break the degeneracy between $B_s$ and $n_e$ in the
 rotation measure (\ref{eqn:rm}), we first consider a method for
 reconstructing electron density fields from possible observed maps of the
 $X$-ray surface brightness and SZ ``flux'', which can be produced
 directly from the simulation data of the clusters.
 We expect that typical clusters can be reasonably approximated as spherically
 symmetric bodies. 
 Then, we can readily write down the $X$-ray surface brightness, $S_X$, 
 and the Compton y parameter for the SZ effect, $S_Y$, at a given frequency band:
\begin{eqnarray}
S_X(r') &=& A_X \int_{-\infty}^{\infty} n_e^2(r) \alpha(T_e(r)) ds , 
\label{eqn:xray}\\
S_Y(r') &=& A_Y \int_{-\infty}^{\infty} n_e(r) T_e(r) ds,
\label{eqn:yray}
\end{eqnarray}
 where $s=\sqrt{r^2-r^{\prime 2}}$, 
 $r'$ and $r$ denote the projected separation and spatial
 radius from the cluster center, respectively, 
 and $T_e$ is the electron temperature.
 Although we will use the physical scale $r'$ throughout, 
 $S_X$ and $S_Y$ can be also given by the angular
 separation $\theta$ from the center via the relation $r'=d_A\theta$,
 where $d_A$ is the angular diameter distance to the cluster.  
 Although $A_X$ and $A_Y$ can be computed once the observational
 frequency bands are specified, 
 their explicit values and expressions  are not relevant for the following analysis.
 Here, $\alpha(T_e)$ is the $X$-ray emission coefficient. We focus on the thermal
 bremsstrahlung only for simplicity and set $\alpha\propto \sqrt{T_e}$, 
 although in reality one
 should take into account line emissions as well. 

 Under the spherical approximation, we can employ the Abel's integral 
 to reconstruct 3D electron density fields following the method developed 
 by Yoshikawa \& Suto (1999). Let us briefly summarize the concept. 
 As in eq. (\ref{eqn:xray}) and (\ref{eqn:yray}), 
 observable quantities $f(r')$ for
 clusters are often written as integrals of the corresponding
  three-dimensional quantities $g$ along the line of sight as
\begin{equation} 
f(r')=\int_{- \infty}^{\infty} g ds=2\int_{r'}^{\infty} g  
\frac{rdr}{\sqrt{r^2-r^{\prime 2}}}.
\end{equation}
 If the quantity $g$ 
 depends on the radius $r$ only (i.e. $g=g(r)$),  we can use the Abel's
 integral to find
\begin{eqnarray}
g(r)= \frac{1}{\pi} \int_{r}^\infty  d{r'} \frac{d f(r')}{dr'} 
\frac{r'}{\sqrt{{r'}^2-r^2}} .
 \label{eqn:abel} 
\end{eqnarray}
 Thus, we can use this transformation formula to deproject the quantities in the
 clusters, assuming spherical symmetry.
 Applying the transformation above to $S_X(r')$ and $S_Y(r')$ given by
 eq. (\ref{eqn:xray}) and (\ref{eqn:yray}) yields
\begin{equation}
n^2_e(r)\sqrt{T_e} = \frac{1}{A_X \pi} \int_{r}^{\infty} dr' \frac{d S_X(r')}{dr'} \frac{r'}{\sqrt{r'^2-r^2}},
\end{equation}
and
\begin{equation}
n_e(r) T_e(r) = \frac{1}{A_Y \pi}\int_{r}^{\infty} dr' \frac{d S_Y(r')}{dr'} \frac{r'}{\sqrt{r'^2-r^2}} .
\end{equation}
From the above two equations, the electron density can be reconstructed as
\begin{equation}
n_e(r)= \frac{\left(1/(A_X \pi) \int dS_X(r')r'/\sqrt{r'^2-r^2} \right)^{2/3}}
 {\left( 1/(A_Y \pi) \int dS_Y(r')r'/\sqrt{r'^2-r^2} \right)^{1/3}} .
\label{eqn:recon}
\end{equation}

\begin{figure}[ht]
\epsscale{0.7}
\plottwo{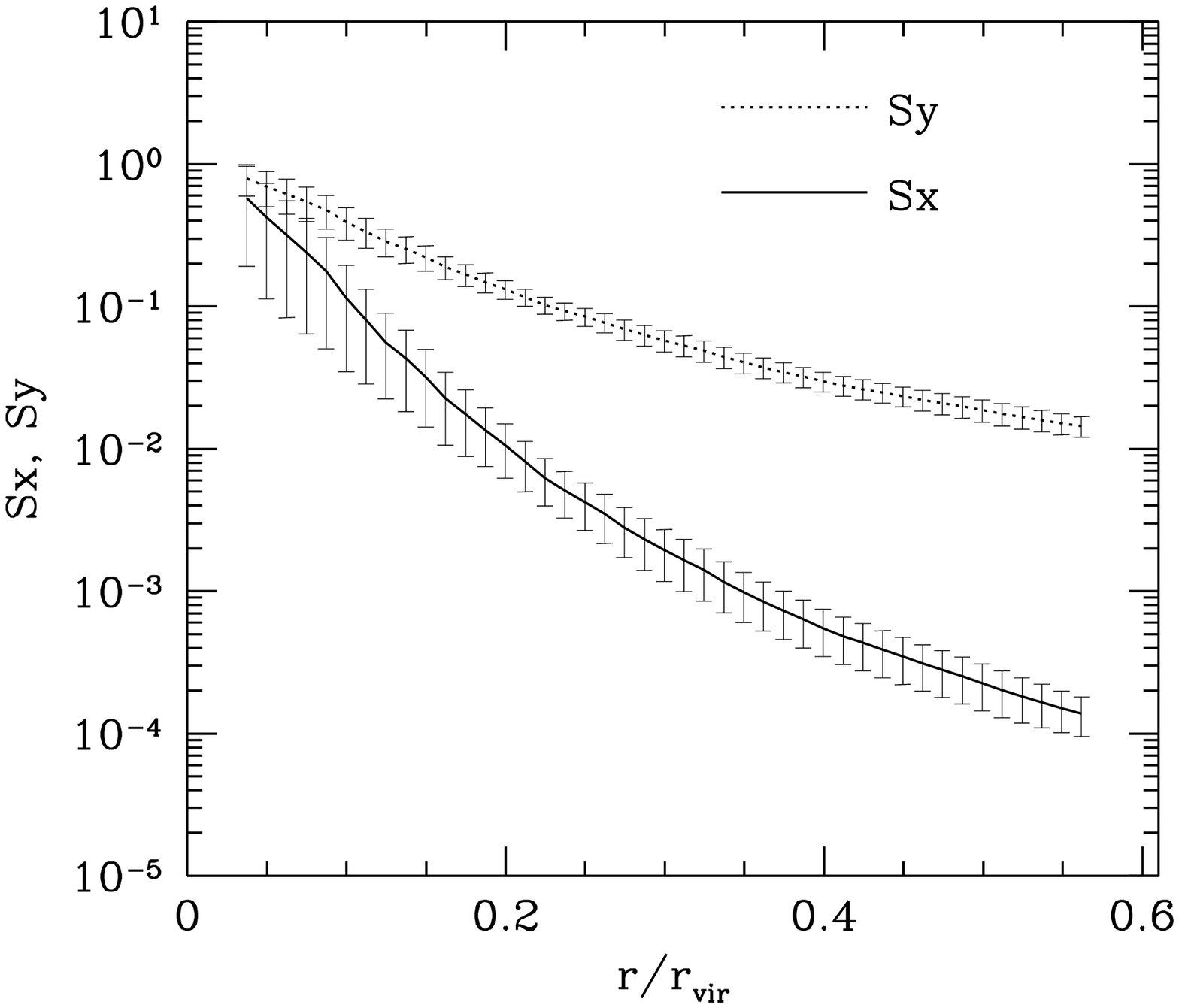}{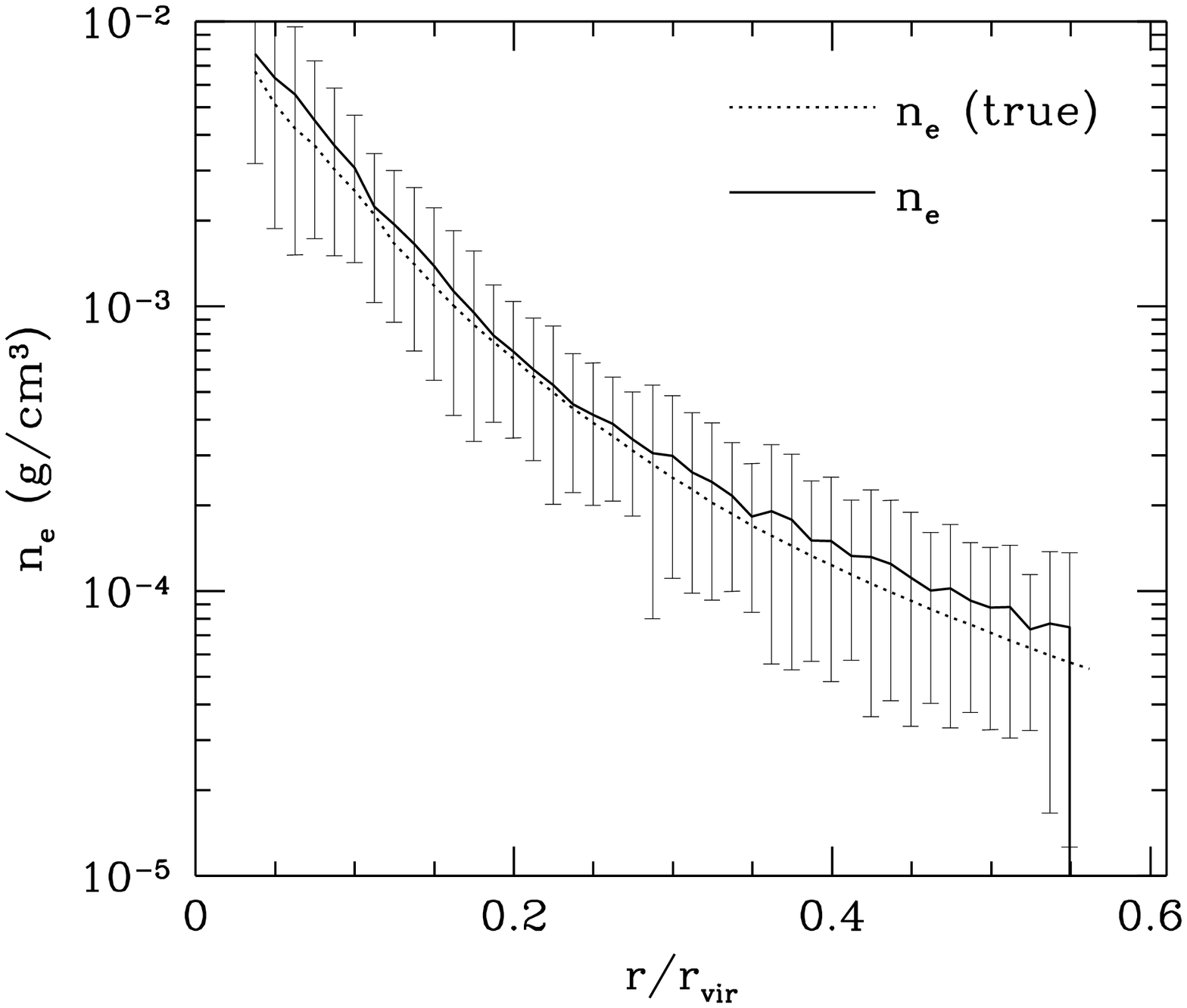}
\caption{In the left panel, the upper line shows the normalized SZ
 $y$-parameter fields, $S_Y$, and  the lower line
 denotes the normalized $X$-ray surface brightness, $S_X$, for 
  cluster model A. 
 The plotted values are the average values of those simulated
 quantities in the annulus of a given projected radius $r'$, while 
 the error bars show the $1 \sigma$ deviation in the same annulus. 
 The horizontal axis
 denotes the radius $r$ normalized by the virial radius $r_{\rm vir}$.
 The right panel shows the reconstructed electron density $n_e$ with
 $\pm 1 \sigma$ errors (see text in more detail) and 
 the true value of spherically averaged density profile  
directly obtained from the simulation.}
\label{fig:re_ne}
\end{figure}

 In the following, we demonstrate the performance of the above
 reconstruction technique.  Since the simulated clusters are not
 spherically symmetric, we cannot directly apply eq.
 (\ref{eqn:recon}) to the simulation data.
 Therefore
 we performed the following procedure along the lines of the Abel's
 integral concept. 
 We first computed the circularly averaged values of the
 simulated $S_X(\bm{r}')$ and $S_Y(\bm{r}')$ in the annulus of a
 given projected radius $r'$. 
 We then employed eq. (\ref{eqn:recon}) to the averaged $S_X$ and
 $S_Y$, giving an estimation of the spherically averaged electron density 
 profile.  The left panel of Fig. \ref{fig:re_ne} 
 shows the averaged values of $S_X(r')$ and $S_Y(r')$ as a function of
 the projected radius $r'$ calculated from the simulated maps of 
 cluster model A,  where the error denotes the 1 $\sigma$ dispersion 
 from the average value in each annulus (see also Yoshikawa \& Suto 1999). 
 Note that we do not attempt to include  observational errors.
 The right panel of 
 Fig. \ref{fig:re_ne} plots the reconstructed electron density profile
 with 1$\sigma $ error bars for cluster A.  For the error
 assignment, we have constructed 200 realizations of the $S_X(r')$ and $S_Y(r')$
 assuming Gaussian distributions of the errors at each bin, and then
 we calculated the average and $1\sigma$ dispersion from the
 reconstructed $n_e$ in the realizations. It is apparent that the
 reconstructed density profile well matches the true spherically averaged
 profile of $n_e$ directly computed from the simulation data. This is
 partly because cluster A model has no distinct substructure and
 the spherical approximation can be valid (see
 Fig. \ref{fig:models}). 

\section{Reconstruction of the Magnetic fields}

 Although one may imagine that the Abel's integral can be similarly applied 
 to the reconstruction of the magnetic fields,
 it is not so easy as the reconstruction of the electron density fields,
 since the magnetic field is a vector quantity and  
 various observations have implied the existence of reversal scale 
 of the intracluster magnetic fields  (e.g., Carilli \& Taylor 2002).   
 The magnetic fields can be treated
 like scalar quantities on scales below the coherence length 
 where the directions of the magnetic fields are the same.
 We
 therefore expect that
 the Abel's integral method can be
 effectively applied in order to estimate a typical field strength for
 the coherence scale. For this reason, we pay attention to developing 
 a method for reconstructing the magnetic field strength as well as a way 
 to estimate the
 coherence length from the observable quantity, i.e., the rotation
 measure. 
 It should be again stressed that the 
 magnetic field strength and the coherence length can be reconstructed
 as a function of radius from the cluster center, which benefits
 from the fact that the CMB polarization field is a continuously varying
 field on the sky, if the secondary effect can be measured.

\subsection{ The dispersion of the rotation measure}

 To apply the Abel's integral to the observed rotation measure, 
 we consider a simplest model of tangled magnetic fields; we assume that  
 the magnetic field strength and the coherence length depend 
 only on the radius from the cluster center.
 This is an assumption of our model, 
 but our results show that it works
 quite well, even despite the complicated
 structure of the coherence length
 within the simulated clusters. 
 Even for this case, we should 
 generally consider the random orientations of the magnetic fields for 
 each cell with the coherence length, 
 whereas the rotation measure along any line of sight can be regarded as a
 random walk process. Hence, the rotation measure in the direction of
 the projected radius $r'$ can 
 be expressed as a sum of the contributions from $N$ cells along the line of
 sight: 
\begin{equation}
\Delta_{RM}(\bm{r}') = \sum_n^{N}   n_{e,n} l_{n} |B_{n}| \cos \chi_n .
\end{equation}
 where  $\chi$ represents the angle 
 between the magnetic field direction and the line of sight. 
% Note that
% the rotation measure field generally depends on the two-dimensional position
% vector $\bm{r}'$ on the sky.
%,  even though $B(r)$ and $l(r)$.  
% To apply the Abel's integral, we next consider
% the rotation measure in the annulus of a given radius range of 
% $r'$ and $r'+dr'$ from the cluster
% center.
% Assuming that the number of cells in the annulus is $M$ and 
 Following the method developed by Lawler \& Dennison (1982), the
 expectation  value of the rotation measure 
 in the annulus is zero and the 1 $\sigma$
 dispersion, $\sigma_{RM}(r')$, can be estimated as
\begin{eqnarray}
{\sigma_{RM}^2  (r') } 
 &=& \sum_n^N \sum_m^M  \langle
  n_{e,n} l_{n} |B_{n}| \cos \chi_n \times 
n_{e,m} l_{m} |B_{m}| \cos \chi_m \rangle,\nonumber \\ 
 &=& \frac{1}{3} \sum_n \left( l_n |B_n| n_{e,}{}_n \right)^2 ,\nonumber\\
 &=& \frac{2}{3} \int_{r'}^{\infty} \frac{r dr}{\sqrt{ {r}^2-r^{\prime 2}}} 
 l(r) {n_e^2(r)} {B^2(r)}, 
\label{eqn:Ibz1}
\end{eqnarray}
 where in the second equation 
 we have used $ \langle \cos^2 \chi_n \rangle = 1/3$, assuming that
 the magnetic field orientations in neighboring spherical shells are
 uncorrelated. 
 Unordered gas motions imprinted during gas formation stir the
 magnetic fields in the simulated clusters well enough 
 for its auto-correlation length to be
 relatively small, of order $50-100 \rm {kpc/h}$.
 This is in good agreement with observations of the Faraday rotation
 patterns (Kim et al. 1990, Feretti et al. 1995).
 Since we can have a sufficient number of cells at a large separation
 $r'$,  eq. (\ref{eqn:Ibz1}) is likely to give a good approximation 
 because of the central limit theorem. 
% The reverse would be also true at
% small separation. 
 Interestingly, eq. (\ref{eqn:Ibz1}) allows us to   
 use the Abel's integral (see eq. (\ref{eqn:abel})) in order to obtain
 the field strength: 
\begin{eqnarray}
 {B^2(r)} = \frac {3}{\pi}  
\frac{1}{l \left( r \right) n_e^2\left( r \right)} \int_{r}^{\infty} 
 \frac {d \sigma_{RM}^2}{d r'} \frac{r'dr'} {\sqrt{r'^2-r^2}}. 
\label{eqn:reco_b}
\end{eqnarray}
This equation means that the magnetic field strength can be estimated
from the rotation measure once the coherence length and the electron
density profile are given. Therefore, the next thing we should do is  
to consider a method for estimating the coherence length from the 
rotation measure fields. 

\subsection{Coherence length of magnetic fields \label{sec:cohe}}

 The two-point correlation function 
 will be a useful and simplest quantity to estimate the coherence
 length of the magnetic fields.  
 Let us first consider the two-point correlation function of the
 magnetic fields, which can be directly computed using
 the simulation data. 
 From the meaning of the rotation measure, we here concentrate on the
 line-of-sight component of the magnetic fields, $B_s$. 
 The the two-point correlation function of $B_s$, $\xi_{B_s}$, can be
 properly defined as
\begin{eqnarray}
\xi_{B_s} (r'|r) \equiv 
\frac {\langle B_s(\bm{x}+\bm{y})  
B_s(\bm{x})\rangle_{{\left|\bm x \right|=r,\left|\bm y \right| =r'}} }
{\langle B_s^2(\bm{x})\rangle}, 
\end{eqnarray}
 where $\bm x$ and $\bm{x}+\bm{y}$ 
 represent the three-dimensional position vectors from the
 cluster center and $\langle\dots\rangle$ denotes the 
 average among all possible pairs within the considered cluster
 subjected to the conditions of $|\bm{x}|=r$ and $|\bm{y}|=r'$. 
 Note that, from our definition of $\xi$, we have
 $\xi_{B_s}(r'=0|r)=1$. 
 Around a radius $r$ from the cluster center, 
 the two points, which are separated by more than the coherence length
 $l$, are considered to be uncorrelated 
 because of the random field orientations, leading to 
 $\xi_{B_s} \approx 0 $ for $r'\simgt l$. 
 Then, 
 the coherence length at a given radius $r$ will be roughly estimated from 
 $r'$ that satisfies the condition $\Delta \xi_{B_s}(r|r') = -1 $. 
 We can thus define the coherence length of magnetic fields as a form of
 the differentiation of the correlation function:
\begin{eqnarray}
 \frac{1} {l(r)} \equiv \left.
\frac{- \Delta \xi(r'|r)}{\Delta r'}\right|_{r'=0}.
\label{eqn:lcoh_defi}
\end{eqnarray}
 Here, to determine $l(r)$ from the simulated clusters, 
 linear fitting is applied 
 where  $\frac{1}{2} \leq \xi \leq 1$ in practice. 

 Likewise, we can define the two-point correlation function of the
 rotation measure fields as
\begin{eqnarray}
\xi_{RM}(r'|r) \equiv
\frac {\langle\Delta_{RM}(\bm{x}+\bm{y})  \Delta_{RM}(\bm{x})
\rangle_{|\bm x|=r,|\bm y|=r'} }
{\langle \Delta_{RM}^2(\bm{x})\rangle},
\end{eqnarray}
 where $\bm x$  represents the projection (2D)
 position vector in a circle of the radius $r$ from the cluster center, 
 and $\bm y$ represents a point in a circle of the
 radius $r'$ from the vector $\bm x$.
% Note that $\bm x$ and $\bm y$ are on the same projected plane because
% the rotation measure is a projected quantity along the line of sight.
 Although the correlation function of the rotation measures contains
 some information on the magnetic field coherence length, we should bear
 in mind that 
 the magnetic field in the rotation measure
 is included not as a single field at a fixed point but 
 as a summation of the fields weighted by the electron density 
 along the line of sight.
 Thus, the two correlation
 functions of the magnetic fields and the rotation measures, $\xi_{B_s}$
 and $\xi_{RM}$, do not
 necessarily match each other. 

\begin{figure}[hb]
\epsscale{0.5}
\plotone{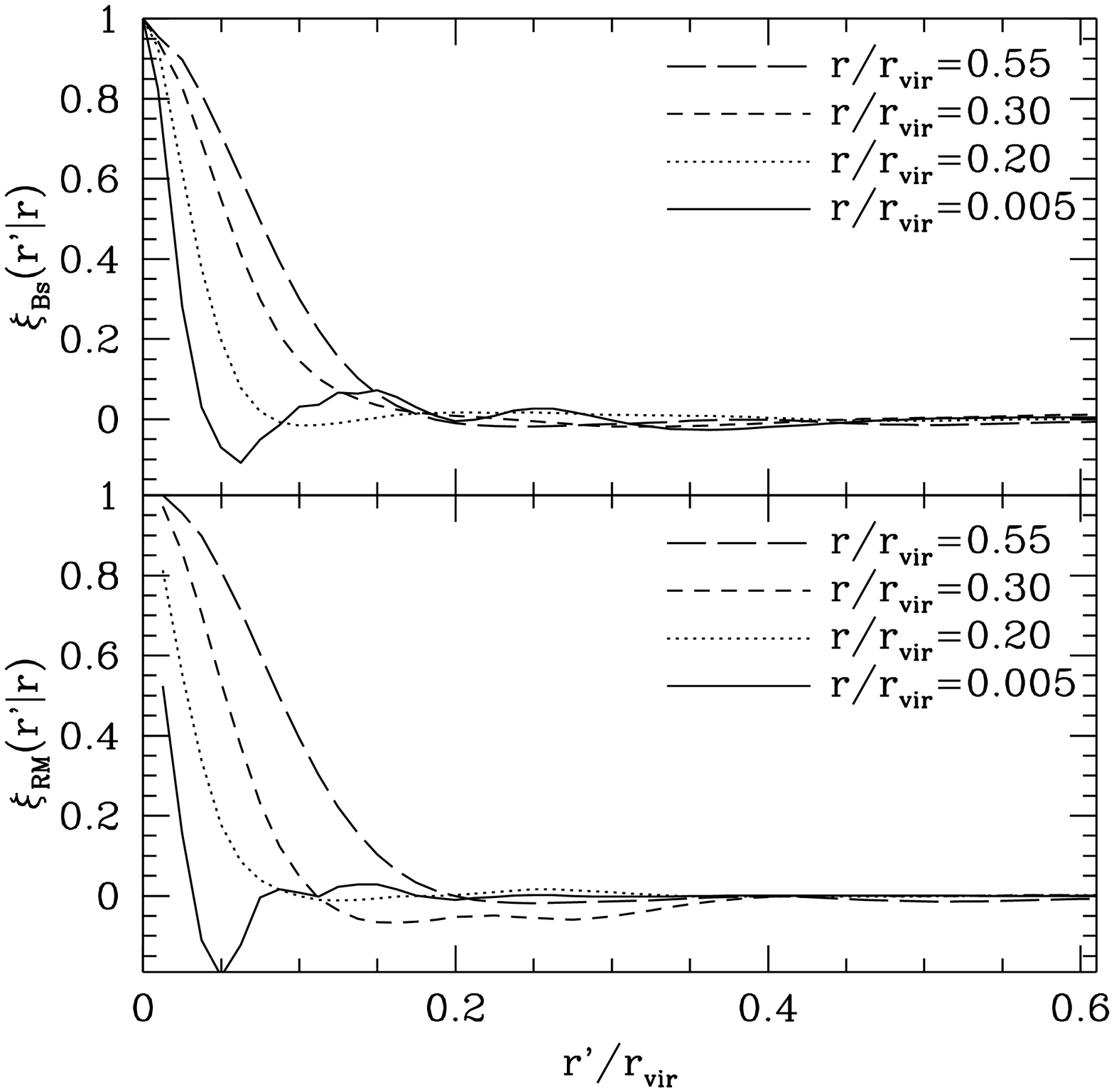}
\caption{The upper panel shows $\xi_{B_s} ( r'|r)$ for cluster A model, and 
 the lower panel shows $\xi_{RM} ( r'|r)$ for the same cluster.
 The correlation functions are plotted for each radius $r$ which is the
 distance from the cluster center.
 The horizontal axis denotes the radius $r'$ normalized by the virial radius
 $r_{\rm vir}$. Note that the radius parameters $r'$ for 
 $\xi_{B_s}$  and $\xi_{RM}$ are distances from a fixed
 point located in a sphere or circle of the radius $r$ 
 from the center of the cluster, respectively.
 }
\label{fig:xi_bzrm}
\epsscale{0.5}
\plotone{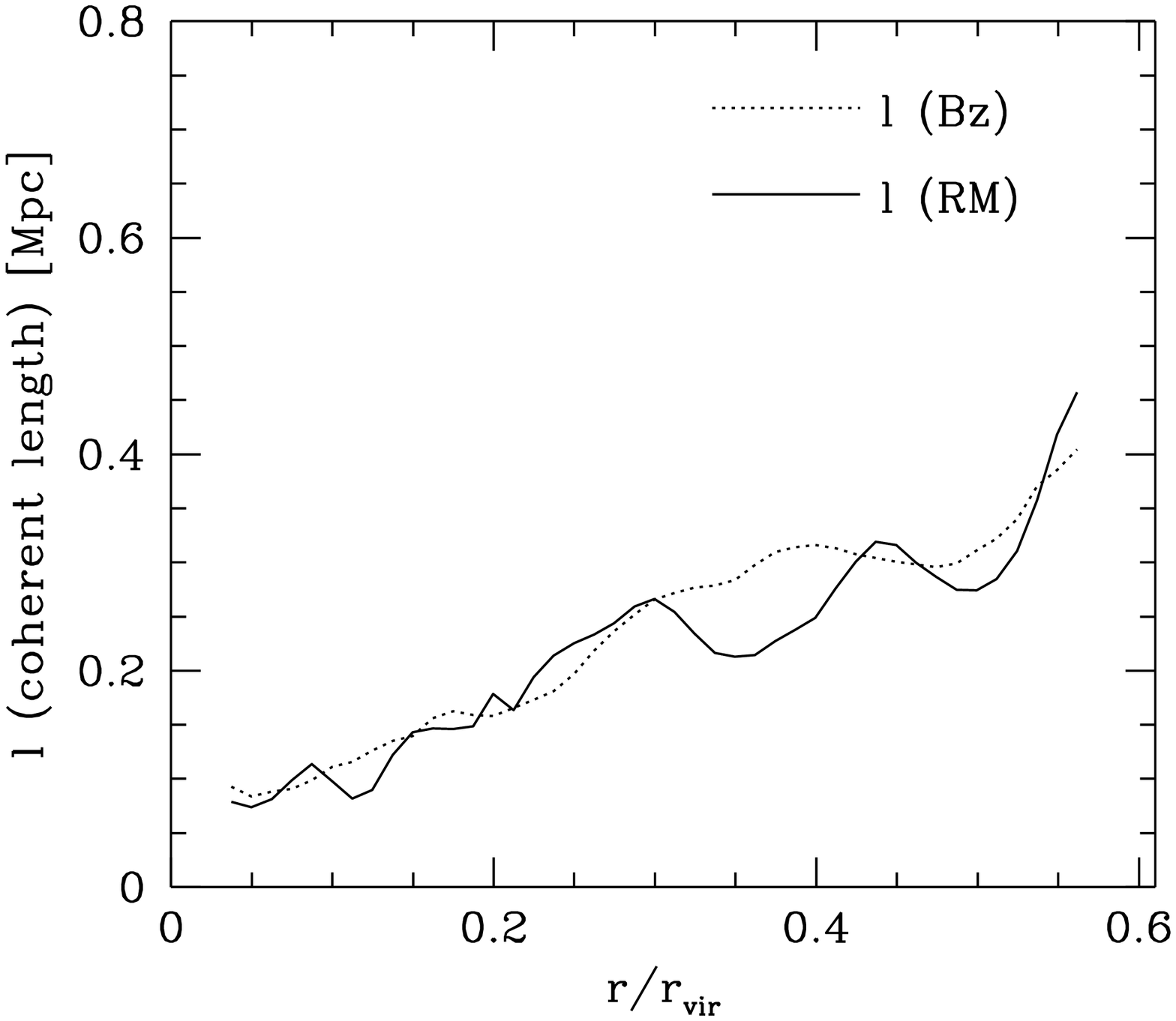}
\caption{The coherence lengths of the magnetic fields in cluster A.
 The horizontal axis denotes the radius $r$ normalized by the virial
 radius $r_{\rm {vir}}$. The dotted and solid curves show the results
 estimated from $\xi_{B_s}$ and $\xi_{RM}$ using
 eq. (\ref{eqn:lcoh_defi}), respectively.  Those coherence lengths well
 match each other within the considered range of radius $r$.}
\label{fig:lco}
\end{figure}

 Fig. \ref{fig:xi_bzrm} plots $\xi_{B_s}$ and $\xi_{RM}$ 
 as a function of $r'$ for cluster A model. We here show the
 results for $r/r_{\rm vir}=0.005$, $0.20$,
 $0.30$ and $0.55$. 
% It is apparent that these correlation functions share the same properties.
% Consequently, the two coherence lengths that are estimated using eq. (\ref{eqn:lcoh_defi})
% from $\xi_{B_s}$ and $\xi_{RM}$ have almost the same features at the
% considered range of $r$ as explicitly shown in Fig. \ref{fig:lco}.
 Although it is not theoretically apparent whether these correlation functions match
 each other, these correlation functions seem to share the same
 properties.
 In fact, as shown in Fig. \ref{fig:lco},
 the two coherence lengths 
 that are estimated using eq. (\ref{eqn:lcoh_defi})
 from $\xi_{B_s}$ and $\xi_{RM}$
 have almost the same features.
 This success can be explained as follows. 
 The rotation measure arises mainly from the region near the cluster center 
 because both the electron density and the magnetic field strength are
 relatively large there in the simulation data.
 Hence, the rotation measure is likely to reflect the property of the
 electron density fields and the magnetic fields on the plane which contains the
 cluster center.
 The magnetic fields determine the sign of the rotation measure because
 of the orientation dependence, 
 and the electron density field is relatively a more smoothly varying
 function with respect to a radius than the magnetic fields.  
 Thus, we could neglect the contribution of the density fields
 to the properties of the two-point correlation function  of the
 rotation measure,  
 and expect that the following relation roughly holds:
\begin{eqnarray}
\xi_{B_s} &\simeq& \xi_{RM}.
\end{eqnarray}
 In fact, we will see in \S \ref{sec:result}
 that this relation is also valid for  
 B and C models of clusters.
 Then, without any additional observational quantities, 
 we can reconstruct the magnetic field strength using the coherence
 length derived from the rotation measure, combined with the reconstructed
 electron density fields (see eq. (\ref{eqn:reco_b})).

\subsection{Results \label{sec:result}}

\begin{figure}[hb]
\epsscale{0.5}
\plotone{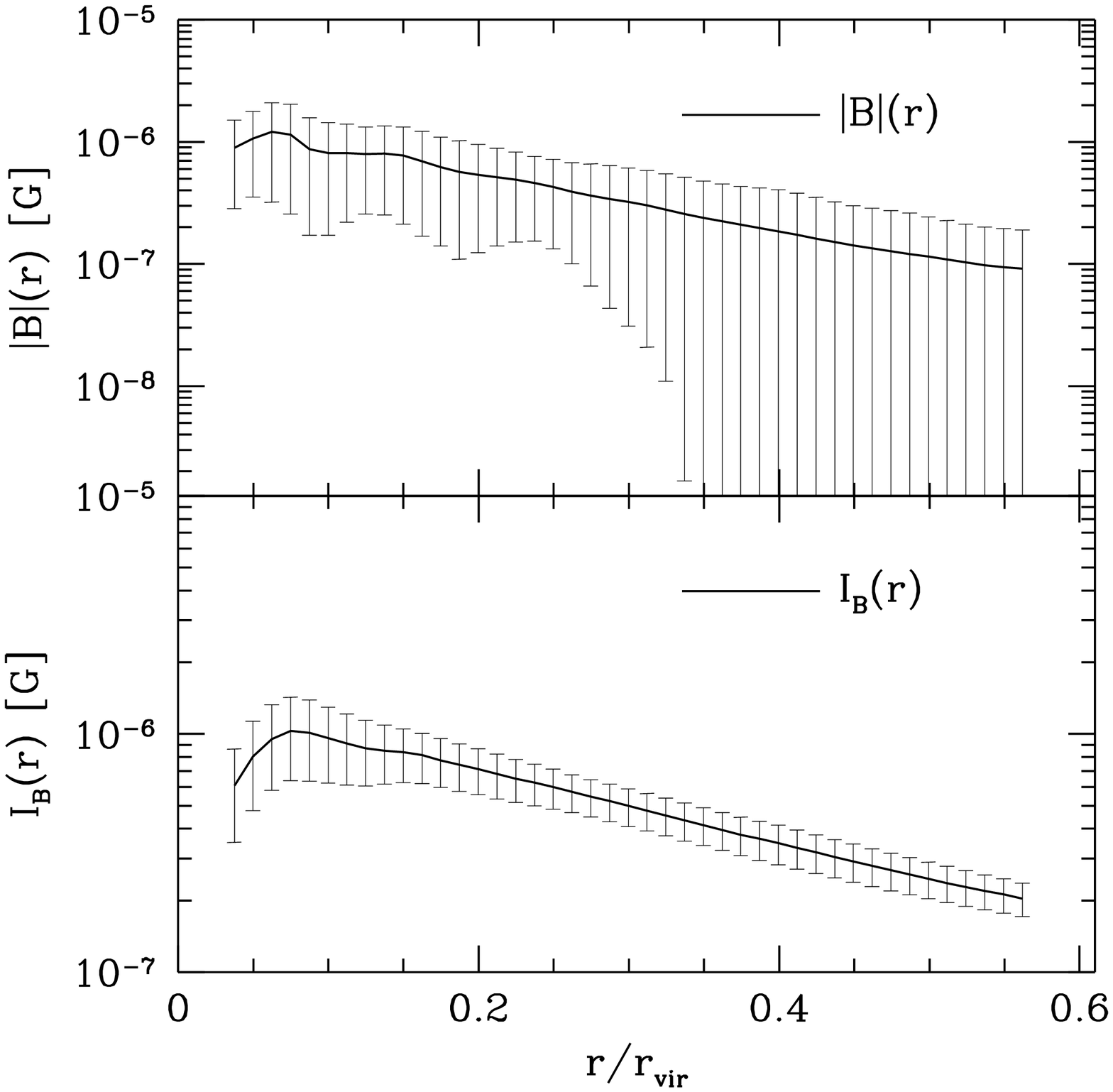}
\caption{The upper panel shows the spherically averaged profile of 
 $|B(\bm{r})|$ for cluster A model, where the error bars denote
 $1\sigma$ deviations from the average value for each bin of 
 radius $r$ as shown in Fig. \ref{fig:re_ne}.
 Similarly, the lower panel shows the cumulative field strength
 $I_B(r) [G]$, which is computed using eq. (\ref{eqn:defib}).
 The horizontal axes denote the radius $r$ normalized by the virial
 radius $r_{\rm {vir}}$.
 }
\label{fig:intBr}
\end{figure}

 We have presented a method for reconstructing the magnetic fields 
 using the Abel integral (\ref{eqn:reco_b}) under a restricted  spherically
 symmetry assumption. 
 In what follows, we demonstrate the performance of 
 our method by comparing the reconstructed results with the true properties of the
 magnetic fields directly calculated from the simulation data.  
 From eq. (\ref{eqn:reco_b}), we expect that our method gives the
 spherically averaged profile of $|B(\bm{r})|$ for a given radius, which
 is shown in the upper panel of Fig. \ref{fig:intBr} for cluster A model 
 as in the left panel of  Fig. \ref{fig:re_ne}. 
 One can see that the averaged profile of
 $B(|\bm{r}|)$ is relatively noisy. This implies that the spherically
 symmetric assumption does not accurately hold for the simulated cluster. 
 For this reason, we will also consider a following cumulative 
 quantity as a radial profile:
\begin{equation}
I_{B}(r) \equiv \frac{\int\!\!  d \Omega  \int_{0}^{r} dr r^2  
\bigl|\bm{B}(\bm{r}) \bigr| } {\int_{0}^{r} dr 4 \pi r^2} .
\label{eqn:defib}
\end{equation}
 Note that, to obtain the reconstructed profile of $I_B(r)$, 
 we first reconstruct $B(r)$ using eq. (\ref{eqn:reco_b}) and then
 compute $I_B(r)$ using the equation above. 
 The profile of $I_B(r)$ for cluster A model 
 can be shown in the lower panel of Fig. \ref{fig:intBr}, which implies 
 that $I_B$ is not so different from $B(r)$ because of its
 outer shell weighted form. 
 The quantity $I_B(r)$ 
 clarifies the tendency of the increase or
 decrease of the magnetic field strength due to  
 its cumulative form.

% Note that the large dispersion of the magnetic field strength may
% increase the dispersion of the rotation measure.
% Then, the field strength which is reconstructed from the
% dispersion of the rotation measure may become larger than the true
% value.
% This effect is not considered in this paper, because it is
% observationally impossible to detect the deviation of the field
% strength from the spherical symmetry.     

\begin{figure}
\epsscale{0.9}
\plotone{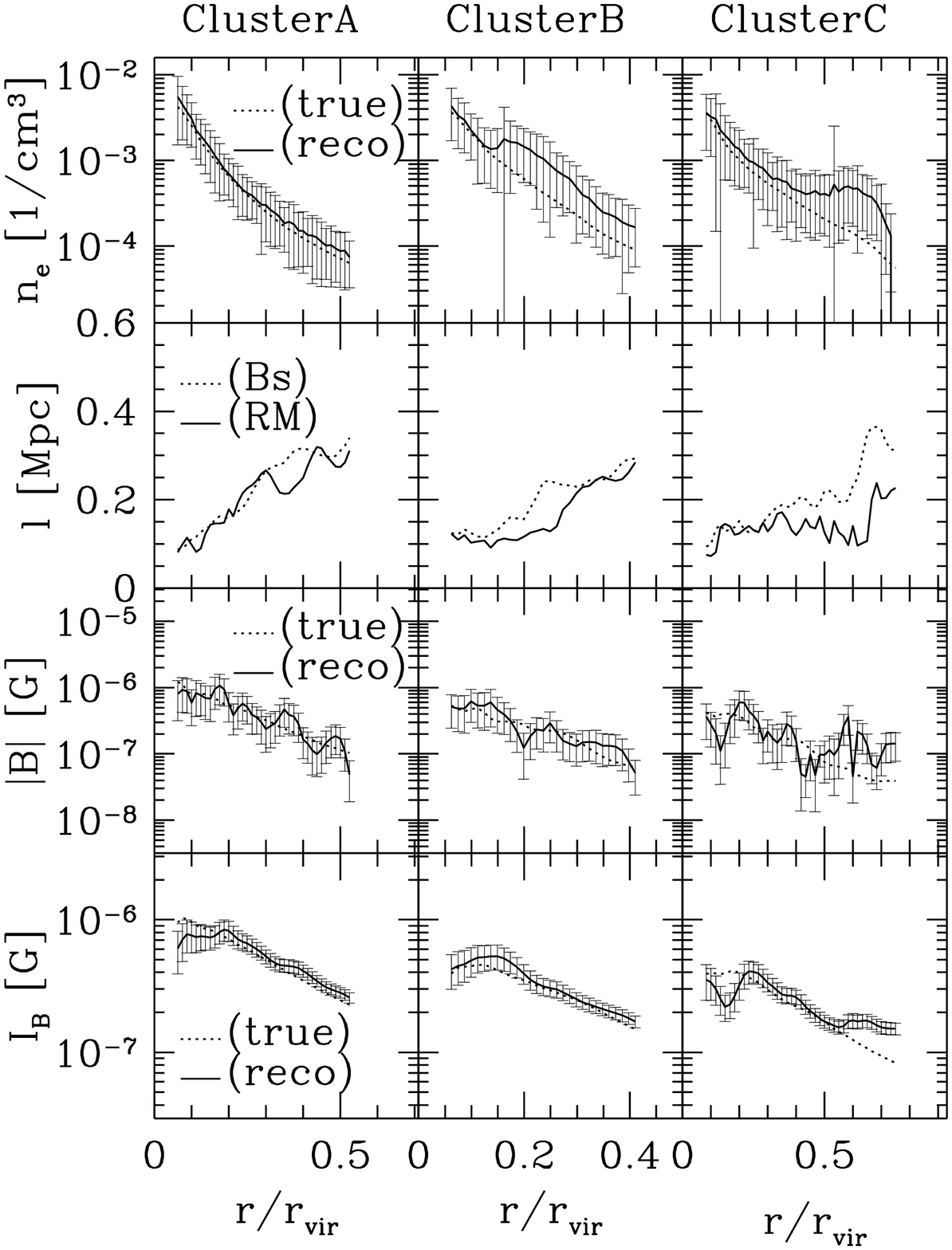}
\caption{These panels show the comparisons of 
 the reconstructed profiles of the electron densities, the coherence
 lengths, the magnetic
 field strengths and the cumulative field strengths with the true profiles
 computed from the simulation data. The left, middle and right panels 
 are the results for clusters A, B, and C, respectively. 
 Those quantities are reconstructed using equations
 (\ref{eqn:recon}), (\ref{eqn:lcoh_defi}), (\ref{eqn:reco_b}) and
 (\ref{eqn:defib}), respectively. 
}
\label{fig:results}
\end{figure}

 Fig. \ref{fig:results}
 shows the reconstructed electron density fields, coherence lengths, 
 magnetic field strengths and their cumulative quantities ($I_B$)
 for cluster models A, B and C.

 Cluster A model is sufficiently virialized and we can consider
 spherical approximation as a reasonable assumption.
 As expected, 
 all quantities of this cluster are well reconstructed.
 One can also see that 
 the coherence length increases with radius $r$ and the coherence
 length at the cluster center is small due to the large random motion of
 the fluid. 
 The coherence length turns out to be an important property of
 the magnetic fields for reconstructing the magnetic field strength
 precisely from the rotation measure.  

 Cluster B model  has two density peaks, 
 and thus should be less virialized than the model A.
 Around the second peak ($r/r_{\rm {vir}} \sim 0.15 $) of this cluster,
 the electron density is reconstructed larger than the true value
 directly calculated from the simulation data,
 because the circularly averaged $X$-ray and SZ fields 
 in the annulus including the second density peak tend to give
 a reconstructed electron density larger than
 the spherically averaged density fields around the radius of the
 second peak (see eq. (\ref{eqn:recon})).
 The coherence length of the rotation measure  relatively matches 
 the value of the magnetic field computed from the simulation data. 
 Only around the second density peak where the
 random motion is large, however,  
 the coherence length is reconstructed small
 relative to the true value. 
 This may be because the random motion effect in the subcluster
 decreases the coherence length of the circularly
 averaged rotation measures more effectively than 
 that of the spherically averaged magnetic fields.
%  However, it is interesting that the profile of 
% the magnetic field strength is well 
% reconstructed as a result of the fact that 
% the overestimation of the reconstructed electron density is compensated
% by the underestimation of the coherence length (see eq. 
% (\ref{eqn:reco_b})).   
 It should be noticed that  
 the increase of the coherence length with radius is weaker than
 in cluster A.  This may be due to the fact that in
 cluster B, which is less virialized than cluster A,
 the magnetic fields are not so efficiently
 tangled by the shear flow as in cluster A.
  Although the magnetic fields are reconstructed with the 
 density fields which deviates from the true value,
 the magnetic field strength seems to be well reconstructed.
 This is because the rotation measures which include
 the projected density fields show the same increase as the reconstructed
 density fields around the density peak, and the magnetic field strength is
 constructed from the ratio of these two values, which cancels out the
 increase of the density fields due to the second peak.

 Cluster C model has a subcluster in the outskirts, since the cluster 
 is at the highest redshift and dynamically youngest system
 than the other two cluster models (see Fig. \ref{fig:models} and
 Table \ref{table:virial}).
 In this cluster,  
 the subcluster raises the reconstructed electron
 density near $r/r_{\rm {vir}} \sim 0.7$ as the result for cluster B
 around the second peak.    
 The coherence length is reconstructed smaller than the true value
 around the subcluster, because the assumption of spherical symmetry
 is not appropriate as the result for cluster B model.
% and the random motion effect in the subcluster should be more effectively
% reflected in the circularly average rotation measures
% than in the spherically averaged magnetic fields.
 We again stress that  the relatively small slope of the coherence 
 length with radius may be due to the insufficient virialization. 
 In this case, the substructure in the outer region
 leads to the result that the magnetic field strength is reconstructed
 larger than the true value. 
% The magnetic field strength also becomes large around the substructure
% due to the large dispersion of the rotation measure which is caused by
% the large deviation of the magnetic field strength from the spherical
% symmetry.
 This overestimation of the magnetic field strength is also caused by
 the deviation from the spherical symmetry due to the concentration of
 the magnetic field in the substructure.
 Nevertheless, it is interesting that the 
 magnetic field strength is relatively well reconstructed at the inner
 range of a radius. 
 
 It is worth clarifying why the magnetic field strength in cluster C
 is relatively stronger in the outer region than the other models.
 The reason was discussed in Dolag et al. (2002): 
 The amplification of the relatively weaker magnetic
 fields in cluster outskirts by adiabatic compression and shear flows
 is more efficient than in cluster cores; firstly because the gas flow
 is more radially ordered outside the core, thus amplification by
 shear flows is more efficient; secondly because the Alfv\'en speed is
 lower in weaker fields and thus closer to the velocity differences
 across flows, which also makes shear flow amplification more
 efficient; and thirdly because the stronger fields in cluster cores
 become dynamically important and thus counter-acts further
 amplification. This explains why magnetic field amplification can
 well be more efficient in dynamically more active, younger systems
 like cluster C. 

 In total, the reconstructed profiles of the strength and coherence
 length of the magnetic fields agree
 with  the true values for the three cluster models, in which the 
 two clusters have the substructures in the inner and outer regions, 
 respectively. 
 

\section{Conclusions}
 In this paper we have constructed a new method for 
 reconstructing magnetic fields in galaxy clusters from
 the Faraday rotation effect on the CMB polarization, combined with 
 possible observed maps of the 
 $X$-ray emission  and SZ effect on the CMB.
 Our results imply that the CMB polarization can be
 potentially used to reconstruct detailed radial
 profiles of the coherence length and
 strength  of the magnetic fields.
 It was shown that the coherence length estimated from the rotation
 measure matches that of the magnetic fields.
 Therefore, we do not need any other
 information than the rotation measure for estimating 
 the magnetic field coherence length.
 This coherence length is not only an important quantity for
 determining the magnetic field strength but also could reveal  
 the origin of the initial seed fields. 

 To reconstruct the field strength, 
 we consider the dispersion of the rotation measure fields in the annulus of
 a given projected radius, motivated by the random walk process caused
 by random orientations of the magnetic fields for a cell with 
 the coherence length.
 However, the deviation of the field strength from the
 spherically averaged value may also increase the dispersion of the
 rotation measure, which we do not consider in this paper.
 This increase of the dispersion
 should lead to the increase of the
 reconstructed magnetic field strength. 
 Nevertheless, the field strengths are well reconstructed in the three
 cluster models, and are reconstructed  within a
 factor of a few even around the area where the dispersion 
 of the field strength is large.  
 Anyway, we expect that  the method constructed in this paper 
 will be a powerful tool for probing the
 intracluster magnetic fields.

 Finally, we comment on the feasibility of this method. 
 It is a great challenge for current technology to detect
 the secondary effect of the intracluster magnetic fields 
 on the CMB polarization. 
 The rotation angle becomes about $1-10^\circ$ at a frequency of 
 $10$GHz (e.g. $\Delta_{RM} \sim 1-10 {}^{o} (10 \rm {GHz}/ \nu)^2$ ). 
 The sensitivity of the detector, which is needed to detect the Faraday
 rotation,  is of order $1\,\mu K$.
 The angular resolution needed to reconstruct the structure of the
 magnetic fields is estimated 
 from the minimum coherence length of the magnetic fields in the
 simulation cluster as $\sim 50 \rm{kpc} \sim 20''$ (at z $\sim 0.1$). 
 The frequency dependence of the Faraday rotation can be also used to
 discriminate the effect from other secondary signals.
 Many observations are ongoing and planned for measuring the CMB polarization.
 We expect that future extensive observations of the CMB polarizations 
 will allow reconstructions of
 intracluster magnetic fields with sufficient accuracy, which should
 give a crucial key to understanding the origin of intracluster 
 magnetic fields.
% This study will be then a crucial key to
%  understand the structures of the magnetic
% fields and possibly their origin. 

 \begin{acknowledgements}
 We thank U. Seljak and M. Zaldarriaga for their available CMBFAST code.
 N.S. is supported by Japanese Grant-in-Aid  for 
 Science Research Fund of the Ministry of Education, No. 14540290.
 N.S. also acknowledges the people in  Institut
 d'Astrophysique Spatiale, Universit\'e Paris-Sud for their 
 kind hospitality.
 \end{acknowledgements}

\end{document}